# Detecting QT prolongation From a Single-lead ECG With Deep Learning

Ridwan Alam PhD[a,b,c,d], Aaron D. Aguirre MD PhD[d,e,f,g], and Collin M. Stultz MD PhD[a,b,c,e,g,h,*]

**Total word count:** 4645
(Text from introduction to conclusions, including references and figure legends)

[a] Research Laboratory of Electronics, MIT, Cambridge, MA, USA
[b] Computer Science & Artificial Intelligence Laboratory, MIT, Cambridge, MA, USA
[c] Department of Electrical Engineering and Computer Science, MIT, Cambridge, MA, USA
[d] Wellman Center for Photomedicine, Massachusetts General Hospital, Boston, MA, USA
[e] Division of Cardiology, Massachusetts General Hospital, Boston, MA, USA
[f] Harvard Medical School, Boston, MA, USA
[g] Harvard-MIT Program in Health Sciences and Technology, Cambridge, MA, USA
[h] Institute for Medical Engineering and Science, MIT, Cambridge, MA, USA

**Funding:** No funding source to disclose.

**Disclosures:** None

**Address for correspondence:**
Collin M. Stultz
MIT, Building 36-796
77 Massachusetts Ave.
Cambridge, MA. 02139
Tel: 617-253-4961
Fax: 617-324-3644
Email: cmstultz@mit.edu

## Abstract

*Background and Aims*
For a number of antiarrhythmics, drug loading requires a 3-day hospitalization with monitoring for QT-prolongation. Automated QT monitoring with wearable ECG monitors would facilitate out-of-hospital care. We aim to develop a deep learning model that infers QT intervals from ECG lead-I – the lead most often acquired from ambulatory ECG monitors – and we use this model to detect clinically meaningful QT-prolongation episodes during Dofetilide drug loading.

*Methods*
Using 4.22 million 12-lead ECG recordings from 903.6 thousand patients at the Massachusetts General Hospital, we develop a deep learning model, QTNet, that infers QT intervals from lead-I. Over 3 million ECGs from 653 thousand patients are used to train the model and an internal-test set containing 633 thousand ECGs from 135 thousand patients was used for testing. QTNet is further evaluated on an external-validation set containing 3.1 million ECGs from 667 thousand patients at another institution.  QTNet was used to detect Dofetilide-induced QT prolongation in a publicly available database (ECGRDVQ-dataset) containing ECGs from subjects enrolled in a clinical trial evaluating the effects of antiarrhythmic drugs.

*Results*
QTNet achieves mean absolute errors of 12.63ms (internal-test) and 12.30ms (external-validation) for estimating absolute QT intervals.  The associated Pearson correlation coefficients are 0.91 (internal-test) and 0.92 (external-validation).  For the ECGRDVQ-dataset, QTNet detects Dofetilide-induced QTc prolongation with 87% sensitivity and 77% specificity.  The negative predictive value of the model is greater than 95% when the pre-test probability of drug-induced QTc prolongation is below 25%.

*Conclusions*
Drug-induced QT prolongation risk can be tracked from ECG lead-I using deep learning. This research leads the path toward out-of-hospital care using wearable ECG devices for antiarrhythmic therapies.

**Abbreviations**
ECG: electrocardiogram
AF: atrial fibrillation
MAE: mean absolute error

**What's New?**

- Using only Lead-I ECG, a novel deep neural network, QTNet, can estimate the QT intervals that are similar to those generated from the 12-lead ECG by the clinical ECG machines, with a mean absolute error of 12ms and a Pearson correlation coefficient of 0.91.
- The same QTNet, when applied without any fine-tuning on an external population undergoing Dofetilide loading, can identify whether and when clinically critical QT prolongation occurs after the drug loading. In comparison to manual annotations of QT intervals by clinical experts, QTNet achieves 87% sensitivity and 77% specificity.
- QTNet is a novel regression model that can be used on Lead-I ECG streams, potentially from wearable devices at out-of-hospital settings, for health critical applications such as drug-induced QT prolongation tracking.

**Introduction**

In addition to stroke prophylaxis, therapeutic options in patients with atrial fibrillation (AF) involve either rate control or rhythm control (1,2). While rhythm control is preferred in patients who cannot tolerate AF, recent data suggests that some patients may benefit from early interventions that reduce AF burden and maintain normal sinus rhythm. This suggests that the maintenance of normal sinus rhythm may be beneficial in patients who do not have symptoms associated with AF (3-5).

A number of antiarrhythmic drugs have proven efficacious with respect to maintaining normal sinus rhythm and shortening the duration and symptoms associated with AF (6-8). As many of these agents have important side effects, however, the initiation of these agents is typically done under the supervision of trained personnel in an inpatient setting. Indeed, the concern for significant QT prolongation necessitates inpatient observation, continuous telemetry, and review of 12-lead ECGs before and after each drug dose (9). For both Sotalol and Dofetilide, for example, current recommendations include 3 days of inpatient hospitalization, corresponding to the administration of 5 drug doses (10).

The advent of wearable and pocket ECG monitoring systems raises the possibility of outpatient drug loading with monitoring for QT prolongation in carefully selected patients (11,12). Central to the success of this approach is a reliable method for estimating QT intervals from ambulatory ECG recordings. Although manual review of ECG recordings by experienced health care providers remains the gold-standard for estimating QT intervals, this approach is limited by the availability of skilled clinicians who can review ambulatory ECG recordings from wearable and pocket ECG devices in a timely manner (11-14). Inspired by the success of deep learning approaches for predicting a variety of clinically meaningful outcomes from the 12-lead electrocardiogram, several investigators have developed methods for automated QT-interval estimation using wearable or pocket ECG monitors (15,16). For example, in one study a deep learning algorithm was designed to predict cardiologist over-read QTc values from a pocket ECG monitoring system. The ability of the method to predict QTc prolongation was evaluated in 686 patients with genetic heart disease – where half of these patients had long QT syndrome – and QTc prolongation was defined as a QTc>500ms (15). Whereas the overall discriminatory ability of the method was good in this cohort, the ability of the model to detect clinically meaningful changes in the QTc was not evaluated. Case in point, for some medications, the threshold to either discontinue drug loading, or change the dosage, corresponds to a 15% increase in the QT or the QTc from baseline after drug administration, which can happen when the QTc<500ms (17). Hence, an algorithm designed to only detect when the QTc>500ms cannot capture all clinically meaningful occurrences of QT prolongation. In another study, an artificial intelligence algorithm (AI-QTc) was applied to 85 patients with Covid-19 who were receiving hydroxychloroquine−azithromycin (16). Overall, the agreement between predicted QTc intervals and cardiologist over-read values was modest at best, and, more importantly, the ability of the method to detect QTc changes over time was not evaluated.

In the present work, we develop a deep learning model that infers QTc intervals from ECG lead I – the lead that is most often acquired using a wearable or pocket monitor ECG monitor (18). Our objective is not limited to continuous QT monitoring, but we also describe an alarm system, based on our algorithm, that identifies QT prolongation during Dofetilide loading.

**Methods**

**Data acquisition**

We used three datasets to develop and evaluate the model. The first dataset, which we refer to as the MGH-dataset, contains 4,223,689 12-lead ECG recordings, from 903,593 patients at the Massachusetts General Hospital (MGH), acquired between January 1981 and December 2020. PR, QRS and QT intervals for each ECG are also stored with each 12-lead ECG in the dataset, which were automatically measured by the ECG acquisition machine (mostly from GE and Philips). All machine generated measurements were included in the metadata of the ECG that were signed off by cardiologists, who often verified the measurements and interpretations if seemed necessary. The second dataset, which we refer to as the BWH-dataset, contains 3,171,283 ECG recordings from 667,060 patients who were seen at the Brigham & Women's Hospital (BWH). The ECG and the interval labels have similar acquisition characteristics, similar clinical equipment was used in collecting those data. The third dataset is a publicly available databank, which we refer to ECGRDVQ-dataset from Physionet, and contains ECGs from 22 healthy subjects during anti-arrhythmic drug loading in a randomized, double-blind, 5-period crossover trial designed to compare the effects of QT prolonging drugs (i.e., Ranolazine, Dofetilide, Verapamil, and Quinidine) versus placebo on electrophysiological parameters (see [https://physionet.org/content/ecgrdvq/1.0.0/)](https://physionet.org/content/ecgrdvq/1.0.0/) (19,20). The labels for QT interval and heart rate on this dataset were manually verified by a clinical expert ECG reader, which guarantees the reliability of the labels for the QT prolongation application.

Retrospective analyses for this study were approved by the Institutional Review Board (IRB) at Mass General Brigham (protocol #2020P000132).

**QTNet Training and Evaluation**

We developed a deep learning model, QTNet, which infers QT intervals and heart rates from 10-second Lead-I ECG signals (Fig 1). QTNet is a multi-task regression Convolutional Neural Network model based on a ResNet-18 architecture (see Supplement for details). Data from the MGH-dataset were used to train and test the model and data from the BWH-dataset were used to validate model results on a population distinct from the training set. Lastly, the ECGRDVQ-dataset was used to test whether the model could identify clinically meaningful QT prolongation during Dofetilide loading.

In detail, the *Training* set contains 70% of the MGH-dataset (3.06 million ECGs, arising from 653 thousand patients). A development set or *Dev* set is used to determine when training should end and corresponds to 15% of MGH-dataset (633k ECG (from 135.5 thousand patients), and the remaining15% of the data comprises our *Internal-test* set, which is used to test the model's performance (633 thousand ECGs from 135.5 thousand patients). Given that each patient general has several ECGs, we ensured that all ECGs from a given patient only appeared in one of these three datasets; i.e., the datasets have no overlap with respect to patient data. Moreover, although ECGs in each of these datasets contain all 12 leads, only data from lead-I is used to train and evaluate model performance.

For training, initial model weights were set according to the Kaiming initialization method with random variables from a normal distribution with variance depending on the layer size (21). Training involves minimizing an objective cost function, which in this case corresponds to the mean square error (MSE) between the predicted QT intervals and heart rates and their true values. For training, QT intervals and heart rates are normalized to zero-mean and unit variance distributions (i.e., Z-scored). Similarly, post-processing involved the reverse operation on model estimations to acquire the absolute values using the mean and the variance from the *Training* set.

Back-propagation with an ADAM optimizer was used to minimize the cost function. For controlling the learning rate, we use a step scheduler to decay the rate in half every 3 epochs, starting from 0.01, and a batch size of 512 was used. Early-stopping was used to reduce risk of overfitting risk based on the validation loss. Essentially, training ends when the MSE starts to rise in the *Dev* set.

In addition to the *Internal-test* set, we evaluated the model on an *External-validation* set that contains 3.1M ECG from 667k patients in the BWH-dataset. For error between the estimated QT intervals and the interval labels, we calculate mean absolute error (MAE). These errors quantify the absolute difference between an individual estimation and its corresponding label and summarize over all estimations as the mean of those differences. Pearson's correlation coefficient (Pearson-R) is used for assessing the amount of similarity between the estimated values and the labels.

**Evaluating QT Prolongation Predictions**

We used the ECGRDVQ-dataset, which contains ECG recordings from 22 healthy subjects who participated in a randomized, double-blind, 5-period crossover trial designed to compare the effects of QT prolonging drugs (i.e., Ranolazine, Dofetilide, Verapamil, and Quinidine) versus placebo on electrophysiological parameters (see https://physionet.org/content/ecgrdvq/1.0.0/) (19). For a given subject, three 12-lead ECGs were recorded 30 minutes before drug administration and again at 15 half-hourly/hourly intervals post-administration, and all QT intervals for each 12-lead ECG in this dataset were adjudicated by the same ECG reader (19). For this study we rely on data corresponding to Dofetilide administration as this is the only medication used in this trial that requires inpatient monitoring during drug loading, and compute the QTc using Bazett's formula (17,22):

$$QTc = QT \times \sqrt{HR \text{ (bpm)}/60 \text{ (s/min)}}$$

where the QT interval and heart rate are obtained as outputs from QTNet. We further define a clinically meaningful instance of QTc prolongation is defined as either:

1) The absolute QTc interval is longer than 500 milliseconds (ms),
2) The QTc interval is 15% longer than that the baseline value, which corresponds to the QTc before the administration of any drug.

This definition is consistent with current guidelines for loading several antiarrhythmic medications, include Dofetilide (17).

We use the procedure outlined in Figure 2 to identify clinically meaningful QTc prolongation events. The procedure uses both QTNet output values, as well as the expected errors in the predictions, which is obtained from the training data. Using the gold-standard, physician adjudicated QTc values, we calculated the sensitivity and specific of our algorithm for identifying clinically meaningful instances of QTc prolongation. Sensitivity values are calculated as the true positive rate – the rate of correct identification of QTc prolongation, and specificity as the true negative rate. For a given sensitivity, specificity, and prevalence of QT prolongation, the positive predictive value (PPV) and negative predictive value (NPV) are computed as follows:

$$PPV = \frac{\text{Sensitivity} \times \text{Prevalence}}{\text{Sensitivity} \times \text{Prevalence} + (1 - \text{Specificity}) \times (1 - \text{Prevalence})} \quad 1$$

$$NPV = \frac{\text{Specificity} \times (1 - \text{Prevalence})}{(1 - \text{Sensitivity}) \times \text{Prevalence} + \text{Specificity} \times (1 - \text{Prevalence})} \quad 2$$

where "Prevalence" refers to the prevalence of QTc prolongation instances in a given population.

**QTc estimation with a Baseline Method**

To compare the performance of the QTNet in both QT interval regression and prolongation detection, we used a baseline ECG delineation algorithm for QT interval and heart rate estimation as implemented in the Neurokit2 library in Python (25). This library implements recent algorithms for ECG analysis on an open code repository on GitHub (see https://github.com/neuropsychology/NeuroKit). We use the signal quality measurement, peak detection, and the delineation algorithms from this library in building our baseline method to estimate QT interval from Lead-I ECG signals.

## Results

Characteristics of the three datasets used to train and evaluate the model are listed in Table 1. The Training and Internal-test datasets correspond to in-house registries of ECGs obtained from patients who receive their care at Massachusetts General Hospital (MGH) and the External-validation set contains patients who receive care at the Brigham and Women's Hospital (BWH). The ECGRDVQ-dataset was used to assess the model's ability to detect clinically meaningful QT prolongation during Dofetilide loading. As the latter dataset is comprised of healthy subjects who were enrolled in a randomized trial – as opposed to patients who are drug loaded to treat an underlying arrhythmia – this population is, on average, younger and have a lower resting heart rate relative to the *MGH* and *BWH* datasets (19).

**Estimating QT intervals with QTNet**

A defining characteristic of QTNet is that it estimates the QTc only using data from Lead-I; i.e., the ECG lead most often acquired by pocket and wearable ECG monitoring devices (11-14). Moreover, instead of directly predicting the QTc using a pre-specified formula, QTNet estimates the absolute QT interval in milliseconds in addition to the average heart rate, where both quantities are inferred using data from Lead-I alone (Figure 1). With these data, the QTc can be calculated using a variety of existing methods; e.g., Bazett, Framingham, Fridericia, and Hodges formulas (22-25).

The performance of QTNet with respect to estimating QT intervals and average heart rates is shown in Figure 3. For both the Internal-test set and the External-validation set, the overall predictive performance for both heart rate and QT interval estimation is excellent, with mean absolute errors (MAE) approximately 12ms for absolute QT estimates and 1.2 bpm for heart rate estimates. Corresponding Pearson correlation coefficients are 0.91 for QT estimates, and 0.99 for heart rate estimates. For comparison, we also tested an established automated method for estimating ECG intervals and heart rates, as implemented in the Neurokit2 library (26). That algorithm was performing comparably for heart rate estimation with an MAE of 2 bpm and Pearson coefficient of 0.96 in the Internal-test set and an MAE of 1.8 bpm and correlation coefficient of 0.97 in the External-validation set. But the estimations of QT intervals were relatively poor; the MAE for QT internals using Neurokit2 was 86.5 ms in the Internal-test set and 90.8 ms in the External-validation set with Pearson correlation coefficients of 0.38 and 0.37, respectively.

**Identifying Drug-induced QT Prolongation**

We used QTNet to identify instances of clinically meaningful QT prolongation in the ECGRDVQ-dataset. Results are shown in Table 2. For comparison, we also list results obtained using the QT interval estimation method in Neurokit2. Overall, QTNet achieves a sensitivity of 87% sensitivity and a 77% specificity while detecting 80% of the instances when QT prolongation occurred. These values are considerably larger than those obtained with Neurokit2.

To gauge how the model can be used in practice, we calculated the positive predictive value (PPV) and negative predictive value (NPV) for QTNet detecting drug-induced QT prolongation events. Both the PPV and the NPV can be calculated from the sensitivity, specificity and the prevalence of the drug-induced prolongation events, as outlined in equations 1 and 2 in the Methods section. For these calculations we used the sensitivity and specificity for QT prolongation events shown in Table 2 and the prevalence of clinically meaningful QT prolongation in the ECGRDVQ-dataset is 25%. The resulting PPV is 56% and the NPV is 95%. More generally, Figure 4 plots the PPV and NPV as a function of the population prevalence and therefore depicts model predictive performance in populations with different intrinsic rates of QT prolongation.

**Discussion**

Inpatient administration and monitoring during drug loading remain a mainstay of clinical practice for a number of antiarrhythmic therapies. Since the purpose of such monitoring is mainly geared towards identifying significant episodes of QT prolongation, we explored whether a machine learning algorithm can be used to estimate QT intervals in an automated fashion, and identify episodes of drug-induced QT prolongation. Furthermore, as our goal is to develop a method that can leverage information obtained from wearable and pocket-held devices, we developed a deep-learning algorithm, QTNet, that only uses data from ECG Lead I; i.e., the lead that is commonly acquired in outpatient ECG monitoring devices. The novel contributions of our approach are not only limited to developing a method that estimates QT intervals from a single-lead ECG but also to test the method's ability to detect clinically meaningful QT prolongation for continuous patient monitoring.

In the current state of practice for antiarrhythmics loading, a 15% increase in the QTc (using Bazett's formula) from baseline after Dofetilide administration warrants a decrease in dosage. Consequently, the physicians who prescribe Dofetilide or other drugs are required to monitor the patient for critical QT changes by continuously reviewing and reading their ECG recordings, limiting the potential number of patients an individual provider can attend to ((13,16). Hence, automated alarm systems for critical QT prolongation, such as QTNet, will significantly impact in both reducing provider burden and increasing the care coverage, even beyond inpatient care.

Additionally, clinical ECG machines that use proprietary algorithms to provide automated interval measurements mostly use lead-II and lead-Vs from the 12 leads of an ECG, hence, cannot be applied as-is when only lead-I is available, i.e. in outpatient or ambulatory settings. Similar concerns are true even for intervals measured by human experts (14,15). As cardiac repolarization is a directional process and the 12 leads of an ECG capture different directions of the related electrical activity, manually measured QT can be different across leads showing QT dispersion. Experts often use lead-II, V1, and V5 to measure QT as those leads better capture the end of the T wave. Lead-I ECG is not a primary choice for "accurate" QT measurement. On the other hand, most wearable and portable ECG devices used in outpatient settings only reliably collect the lead-I ECG. Hence, estimating the QT intervals from lead-I ECG, as QTNet accurately performs, is a novel challenging task of significant healthcare utility.

QTNet hypothesizes that the lead-I ECG contains sufficient information about the "accurate" QT interval and uses deep learning to extract and utilize that information. For training and evaluating QTNet in estimating QT intervals, we use the ECG machine generated QT intervals associated with the ECG that was signed off by a cardiologist as the labels or ground truths. These intervals are measured by the proprietary (GE/Phillips) algorithms using all 12 leads. QTNet learns to estimate those QT labels from lead-I ECG only, even though lead-I is not the best electrical axis to capture the cardiac repolarization, but the best suitable option to generalize beyond inpatient care setting. The regression performance demonstrates strong support for that hypothesis and highlights the potential of estimating accurate QT intervals from ambulatory wearable ECG devices. In our holdout *internal-test* data from MGH and *external-validation* data from BWH, the performance by QTNet in estimating QT interval remains reliably consistent with mean absolute errors of 12.63ms and 12.30, respectively. Comparing with a recent study, the median absolute error reported between QTc values using lead I of the Apple Watch and the 12-lead ECG is 18ms (13).

We further demonstrate the utility of the QTNet estimated QT intervals in a clinically important application of QT prolongation detection on public dataset *ECGRDVQ* from Physionet (20). QTNet can identify instances of Dofetilide-induced QT prolongation that were identified from an analysis of the corresponding 12-lead ECG. In this dataset, the labels for the QT intervals are algorithmically annotated and manually verified by a clinical expert, hence, the dataset closely represents the clinical application setting (19). Using QTNet, most notably, the associated sensitivity and specificity for identifying drug-induced QT-prolongation was 0.87 and 0.77, respectively. The corresponding NPV was 95% in this cohort, which has an underlying prevalence of 25% for QT prolongation events.

Patients who are deemed to be at high risk of drug-induced QT prolongation (e.g., individuals with an elevated QTc at baseline or who are already taking medications associated with QT-prolongation, etc.) benefit from inpatient drug loading – in keeping with current guidelines. However, it is not clear whether low risk patients (e.g., normal QTc at baseline, no concomitant use of QT-prolonging drugs, no family history of QT-prolongation, etc.), derive the same benefit.

Indeed, outpatient drug-load with contemporaneous QT monitoring via a wearable or pocket ECG monitoring device may be viable mechanism for patients who have a low pre-test probability of drug-induced QT-prolongation. Our results suggest that QTNet has a high NPV for identifying instances of QT-prolongation when the underlying population has a low prevalence of QT-prolongation (Figure 4). As patients who are deemed to be low risk, by definition, comprise a population with a low prevalence for QT-prolongation events, the NPV of QTNet is expected to be similarly large when the pre-test probability is low. Therefore, a negative prediction in patients who have a low pre-test probability for drug-induced QT-prolongation strongly suggests that clinically significant QT prolongation is not present and that drug dosing can continue without modifying the drug dose.

**Study Limitations**

Our method was developed to use signals corresponding to ECG Lead-I, which corresponds to the signal often acquired in wearable outpatient ECG monitors. As we do not explicitly use data arising from such wearable devices, however, our approach requires further prospective validation in an ambulatory setting. QTNet associated sensitivity and specificity for identifying instances of clinically meaningful QTc prolongation events were calculated from a population of healthy individuals who were given Dofetilide. Further prospective validation in low-risk clinical cohorts is warranted. Similarly, the general applicability of QTNet across ethnicity, race, and age is yet to be explored.

**Conclusions**

The QT interval and heart rate can be accurately estimated from a single lead ECG using deep learning. Clinically meaningful instances of QT prolongation can further be identified using a single lead, thereby enabling outpatient monitoring for QT prolongation in patients who have a low-pretest probability for drug-induced QT prolongation.

**Data sharing**

Model weights and architecture are available at http://github.com/mit-ccrg/QTNet, and the data on drug-induced QT prolongation are available online at http://physionet.org/content/ecgrdvq.

**Clinical Perspectives**

*Competency in Systems Based Practice:* Deep learning can be leveraged to estimate when the QT interval is prolonged using ECG Lead I, which is the ECG lead most often acquired by wearable devices.

*Translational Outlook:* Ambulatory ECG monitoring, coupled with sophisticated algorithms for identifying instances of QT prolongation, can enable outpatient Dofetilide-loading in low-risk patients

**Tables**

*Table 1:* Demographics of patient population and other descriptions across the three datasets.

| Dataset | Training | Internal-test | External-validation | ECGRDVQ |
|---|---|---|---|---|
| Number of Patients | 652,845 | 135,539 | 667,060 | 22 |
| Total number of ECGs | 3,056,660 | 632,997 | 3,171,283 | 5232 |
| Age (yr) | 60.6 ± 18.7 | 60.4 ± 18.7 | 60.3 ± 16.4 | 27 ± 5.4 |
| Female (%) | 42.8 | 43.3 | 50.2 | 49.5 |
| HR (bpm) | 77 ± 20.3 | 77 ± 20.2 | 77 ± 18.7 | 64 ± 9.5 |
| QT (ms) | 394 ± 50.0 | 394 ± 49.8 | 396 ± 47.6 | 400 ± 33.7 |
| QTc (ms) | 438 ± 38.5 | 438 ± 38.3 | 440 ± 35.3 | 412 ± 36.2 |

*Table 2:* Detecting QT prolongation instances after Dofetilide dosing on the RDVQ-dataset; the prevalence for the prolonged QT interval is 25% from the 330 temporal instances.

| **Methods** | **Neurokit2** | **QTNet** |
|---|---|---|
| Sensitivity | 0.29 | 0.87 |
| Specificity | 0.83 | 0.77 |
| Accuracy | 0.70 | 0.80 |

# Figures

*Representative Figure* : **QTNet tracks the QT prolongation risk**. QTNet estimates QT intervals from 10-second Lead-I ECG signal, which are used to assess whether QT is prolonged at any instance. In this example plot, Dofetilide 500ug was administered at time t=0 hour for a patient, and the estimated values of QT interval and the corresponding labels are shown across time. The estimations by QTNet follow the labels within one MAE prediction bounds. QT prolongation instances were detected at t=1.5 and t=2 comparing the estimations at those instances with the pre-dosing estimation (at t=-0.5 hour). The predicted alarm and the label for the QT prolongation overlap at t=2 (two hours after administration).

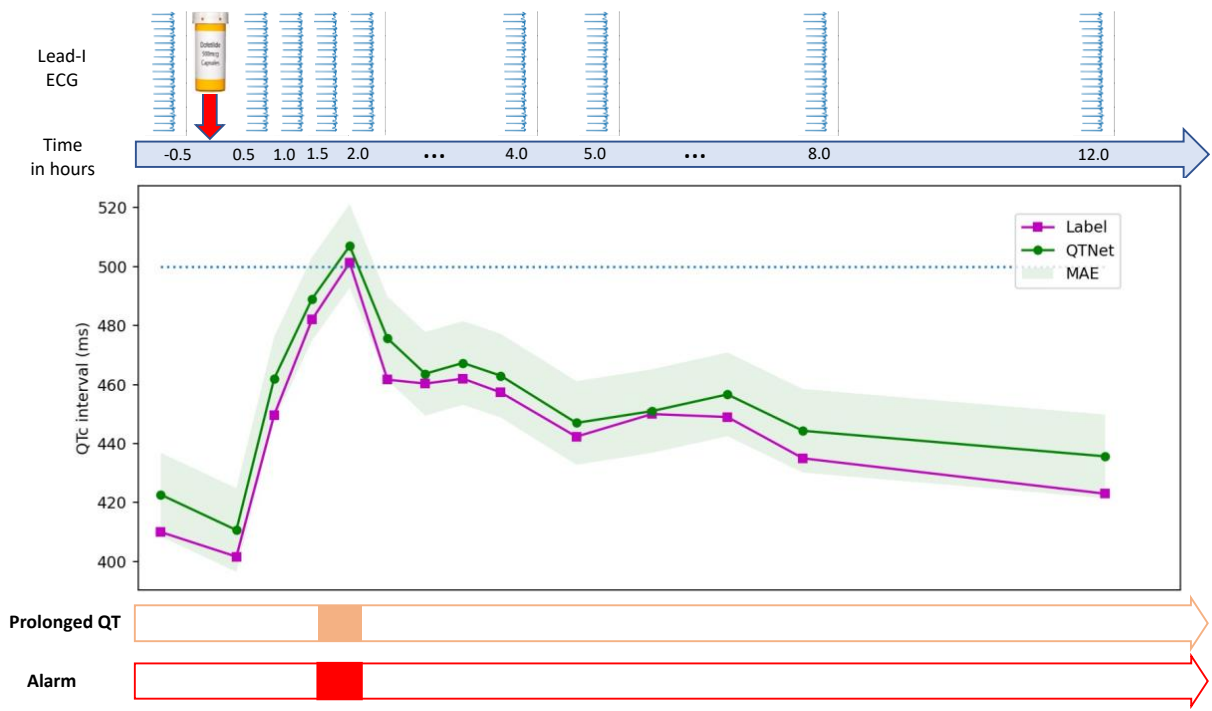

*Figure 1:* **QTNet regression model.** The model is trained on Lead-I 10-second ECG signals to estimate both the QT intervals and the heart rates.

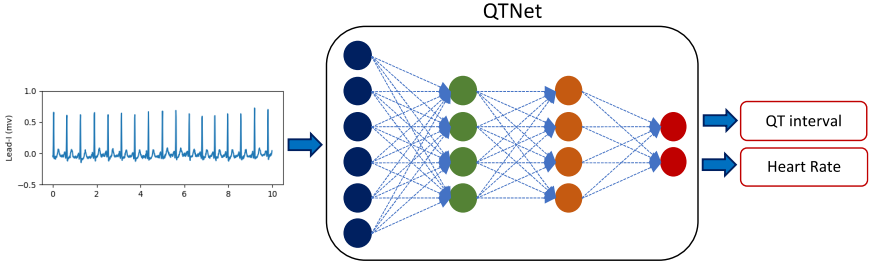

*Figure 2:* **QT prolongation detection from QTNet.** At any time t, QTNet is used to estimate the QT interval and the heart rate, which are then used to calculate the corrected QT interval using Bazett's formula and adjust for the distribution shift using *Training*-MAE. Half-hour before drug loading (here, t = 0), we calculate adjusted QTc(0). At regular intervals t=T after the drug administration, the QTc(T) is calculated and adjusted, and used to detect possible prolongation. The detection is triggered if the adjusted value is greater than 500ms, or, if QTc(T) is increased by 15% compared to the pre-dosing adjusted interval QTc(0).

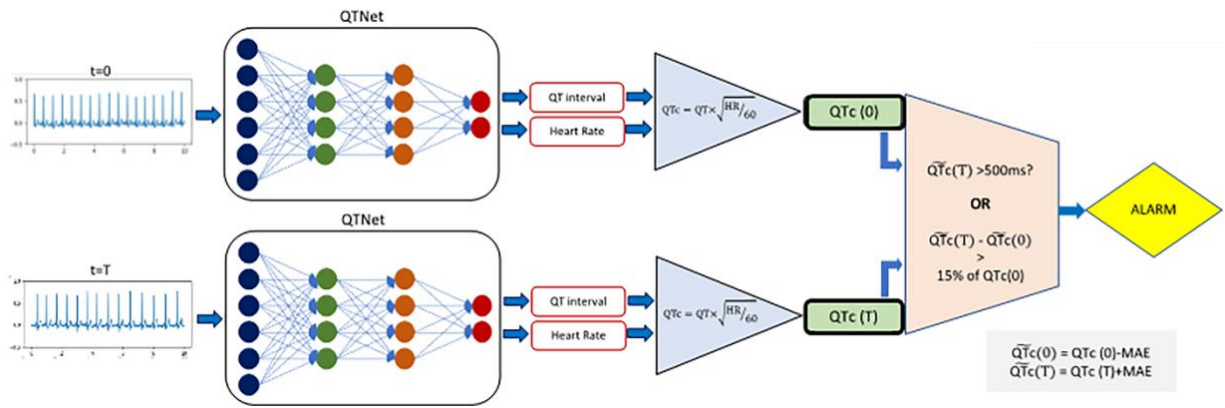

*Figure 3:* **QTNet regression performance on test data.** The estimated QT intervals and HR are compared with the corresponding labels on two datasets: *Internal-test* (MGH test) and *External-validation* (BWH) using two metrics of performance: mean absolute error (MAE) and Pearson-R correlation coefficient. The plots show the comparative performance of QTNet with the baseline algorithm implemented using Neurokit2 library. (a) The estimation errors for QT estimation using QTNet are very low (about 12 milliseconds) and the correlation coefficients are very high (0.91) on the *Internal-test* set, and (b) remains steady even to *External-validation* set which was not seen by QTNet during training. (c) and (d) show notable performances in HR estimation both by QTNet and the Neurokit algorithm across datasets.

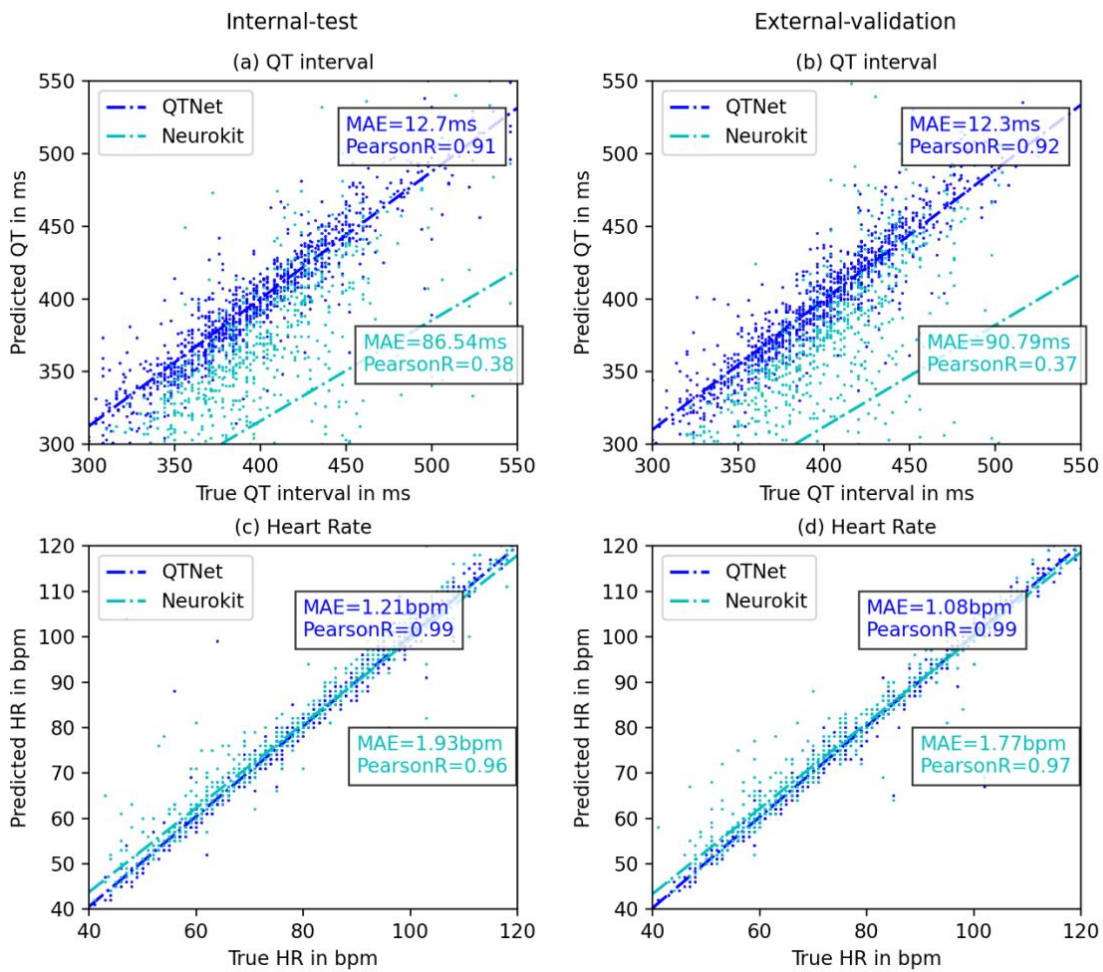

*Figure 4:* **Predictive values of QTNet in detecting QT prolongation.** The RDVQ dataset contains 24 hours data (N=330) from 22 subjects. In 25% of those instances, mostly within 3 hours of Dofetilide administration, the QT prolongation was observed. QTNet can detect 80% of those instances with 87% sensitivity and 77% specificity. Using Equation (1) and (2), the negative and positive predictive values (NPV and PPV) are plotted for varying prevalence.

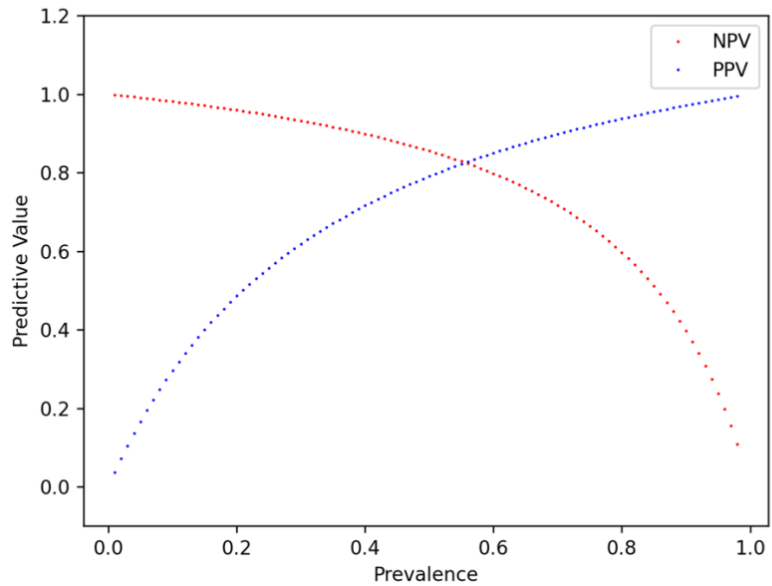